# Dark Self-Healing Mediated Negative Photoconductivity of Lead-Free $Cs_3Bi_2Cl_9$ Perovskite Single Crystal

Naveen Kumar Tailor,[1] Partha Maity,[2] Makhsud Saidaminov,[3] Narayan Pradhan,[4] Soumitra Satapathi[1*]

[1]Department of Physics, Indian Institute of Technology Roorkee, Roorkee, Haridwar, Uttarakhand, 247667, India

[2]Division of Physical Sciences and Engineering, King Abdullah University of Science and Technology (KAUST),Thuwal 23955-6900, Saudi Arabia

[3]Department of Chemistry, University of Victoria, Victoria, BC V8W 2Y2, Canada

[4]Department of Materials Science, Indian Association for the Cultivation of Science, Kolkata, India

**Corresponding Author-** soumitra.satapathi@ph.iitr.ac.in

## Summary

Halide perovskites are recently emerged as one of the frontline optoelectronic materials for device applications and have been extensively studied in past few years. Among these while, lead-based materials were most widely explored, investigation of optical properties of lead-free perovskites is limited. Being optically active, these materials were expected to show light-induced enhanced photoconductivity and the same was reported for lead halide perovskite single crystals. However, on contrary, herein, light-induced degradation of bismuth halide perovskite $Cs_3Bi_2Cl_9$ single crystals is reported which was evidenced by negative photoconductivity with slow recovery. The femtosecond transient reflectance (fs-TR) spectroscopy studies further revealed these electronic transport properties were due to the formation of light-activated metastable trap states within the perovskite crystal. The figure of merits of $Cs_3Bi_2Cl_9$ single-crystal detectors such as responsivity (17 mA/W), detectivity (6.23 × $10^{11}$ Jones) and the ratio of current in dark to light (~7160) was calculated and it is found that they are comparable or higher to reported perovskite single crystals based positive photodetectors. This observation for lead-free perovskite single crystals which were optically active but showed retroactive photocurrent on irradiation remained unique for such materials.

**Introduction**

Halide perovskites have recently drawn significant research interest for their large absorption coefficient, tunable bandgap and long carrier diffusion length.[1-4] These optoelectronic materials remained in forefront of current research for both light-emitting and photovoltaic applications.[5-8] Among all these materials, Pb-based perovskites dominated the entire research because of their faveolus properties, considerably longer stability and also their formation processes that were widely understood.[9-13] From nanocrystal to bulk single crystal and from 0D to 2D and 3D materials, these lead halide perovskite materials remained in the front line.[5, 14-18] However, recent progress has also revealed the success of Pb-free perovskites having acceptable level optical properties.[19-21] Among these, lead-free double perovskites with a group (I) and group (III) metal ions and Sn, Sb and Bi-based alternative perovskite materials were the few attractive candidates reported so far.[17, 22-24] However, extensive research has already been focused on developing such new materials without compromising the properties of the material.[25-28]

Being photoactive, one of the most widely studied properties of these materials is their photoconductivity. For semiconducting nature, these were expected to show enhanced photocurrent or positive photoconductivity (PPC) on irradiation and the same were also reported for nanocrystals as well as some bulk single-crystal materials. Ding et al. showed the PPC in $CsPbBr_3$ single-crystal with high detectivity and rapid response.[29] Makhsud et al. also presented the PPC and self-powered photodetection in $CsPbBr_3$ single crystals.[30] These suggested that in lead halide single crystals positive photoconductivity could be observed with different anion and cation variation with their different response and detectivity.[31-39] Similarly, Bin Yang et al. demonstrated in lead-free $MA_3Sb_2I_9$ single crystals exhibiting PPC with fast response and sensitive detection.[40] Chengmin Ji et al. showed the high performance of lead-free (TMHD)$BiBr_5$ single crystals with positive photoconduction and rapid response.[41] Jun Zhou et al. presented the narrowband photodetection with the positive photoconductivity in lead-free perovskite derivative $Cs_2SnCl_{6-x}Br_x$ single crystals.[42]

However, in some unusual cases, the conductivity also reduced below the dark level under illumination and this was termed as negative photoconductivity (NPC). Although NPC is often considered as anomalous, there have been reports on the NPC phenomenon.[43-50] Intriguingly, herein, negative photoconductivity was also reported for $Cs_3Bi_2Cl_9$ single crystals. The current degradation was observed increased significantly with an illumination of 15 minutes but showed a slow recovery in the dark to its original value. This observation was further confirmed by impedance spectroscopy. The study of transient reflectance spectroscopy revealed that the negative photoconductivity and dark self-recovery might be due to the formation of metastable states in the perovskite, which trapped charge carriers and released slowly in dark. Evidence of phase change or crystal deformation could not be observed as optical features remained similar to the original single crystal material. Bi-based perovskites were largely accepted as lead-free materials and reported with their thermal and moisture stability. Besides, these materials have been demonstrated the high sensitivity for X-ray detectors[51-54]. However, negative photoconductivity here adds a new feature to the electronic properties in the perovskite family of material.

**Results and Discussion**

The lead-free perovskite $Cs_3Bi_2Cl_9$ single crystals were grown following a modified method reported in our previous paper.[55] The slow cooling led to an approximate 4 mm $Cs_3Bi_2Cl_9$ single crystal. Figure 1a presents the digital image of the single crystal and the scanning electron microscope image is shown in Figure S1, which shows a grain-boundary free surface. Figure 1b shows the single-crystal X-ray diffraction pattern of $Cs_3Bi_2Cl_9$ crystal. Details of the analysis of these peaks are provided in the supporting information and from their positions, these were confirmed orthorhombic phase. Crystal information file (CIF) from the obtained single crystal is also provided in supporting information. The atomic model showing the unit cell for the obtained $Cs_3Bi_2Cl_9$ crystal is presented in Figure 1c. The optical absorption spectra (Figure 1d) of this single crystal showed the band edge around 410 nm and the calculated bandgap is shown in Figure S2. The photoluminescence spectra excited at 340 nm is shown in Figure 1e which featured a narrow band and also blue-shifted in comparison to the band edge. Similar blue shifted and narrow emissions were also reported for $CH_3NH_3PbX_3$ (X= Cl, Br, I) single crystals[56] and this characteristic was suggested because of minimized or trap free single crystal. Interestingly, the decay lifetime (Figure 1f) remained in picoseconds indicating atomic or small clusters like transition in these single crystals.

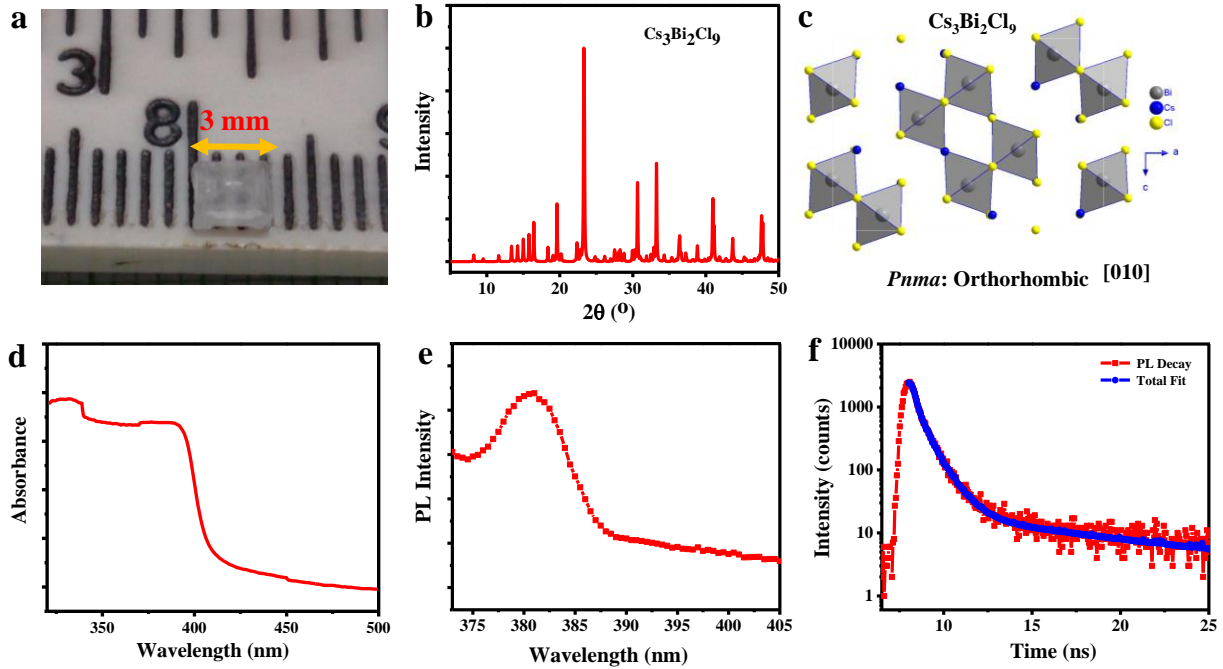

**Figure 1. Structural and Optical Properties**
(a) Digital image of nearly 3 mm single-crystal $Cs_3Bi_2Cl_9$ perovskite.
(b) Single crystal X-ray diffraction of the grown $Cs_3Bi_2Cl_9$ crystal. Assignments of peaks and the CIF file are provided in supporting information.
(c) An atomic model of the unit cell of $Cs_3Bi_2Cl_9$ perovskite was obtained using the obtained CIF data.
(d) Absorption spectra,
(e) Corresponding photoluminescence spectra, and
(f) Excited state decay plot of $Cs_3Bi_2Cl_9$ single crystal. The exciting wavelength is 340 nm and the emission is at 379 nm for the decay plot.

The current-voltage characteristics of the $Cs_3Bi_2Cl_9$ single crystal device are measured under 1-sun illumination (Air Mass 1.5) in the planar device structure as shown in Figure 2a. Figure 2b shows the typical I-V characteristics with different light exposure time and corresponding self-recovery again in dark. As observed, the dark current starts decreasing when the light incident on the single crystal. Current decreases completely after 15 minutes of constant 1-Sun illumination and after then current recovers to its original steady-state value in 180 minutes after resting the device in the dark (Figure 2c). The current below the dark level demonstrates the negative photoconductivity in the $Cs_3Bi_2Cl_9$ crystal.

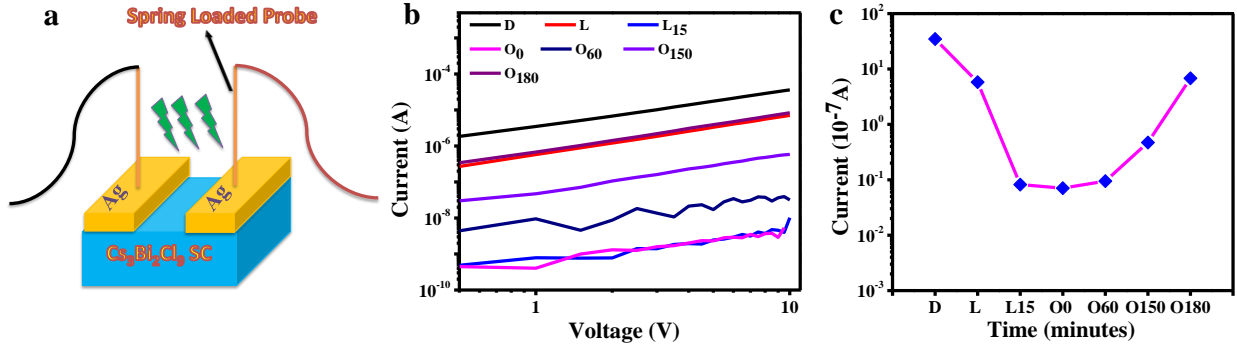

**Figure 2. Device Structure and I-V Characteristics**
(a) The device structure of $Cs_3Bi_2Cl_9$ single crystal, which is used for the electrical measurements.
(b) Current-Voltage characterization of $Cs_3Bi_2Cl_9$ single crystal under 1-sun illumination. Photocurrent profile under different applied voltages as a function of device light-illumination time and dark-storage time. It shows photocurrent degradation of the device under constant light-illumination and self-recovery of the photocurrent in the dark.
(c) Current (at 1 V) vs time curve in dark and light illumination conditions. Here D represents the dark condition. L represents the light illumination condition. O represents light off condition.

To confirm the observations in Figure 2, we also perform the impedance spectroscopy (IS) in $Cs_3Bi_2Cl_9$ single crystal. Figure 3a illustrates the representative Nyquist (Z'-Z'') plots of impedance spectra of the $Cs_3Bi_2Cl_9$ perovskite single-crystal device in dark, with light exposure and dark recovery respectively. We observe that the diameter of the semicircle increases with in illumination condition, which indicates the enhancement of the resistance ($R(\omega) = Re(Z)$) in consistent with the trend observed in the bode plots (Figure 3b). The enhancement of the resistance is correlated to the reduction of the capacitance in the low-frequency region.[57] After light off, the diameter of the semicircle decreases and comes back to dark conditions. The equivalent circuit is depicted in the inset (Figure 3a) in which a constant phase element with a parallel resistor element is added with series resistance. The EIS data was fitted using ZSimpWin Software and obtained the element values (Table S1). The impedance of the device increases with light exposure time, which is contrary to the expected observation that light exposure impedance should decrease as the number of photo-generated current increases with light excitation. The behavior of real and imaginary components of impedance with light exposure and recovery time is displayed in Figure S5. This unusual photo-impedance observation can be modeled with a two-terminal device, which consists of $Cs_3Bi_2Cl_9$ perovskite as the active layer and metal electrode as the contact. If the perovskite-electrode contact is perfectly reflective then due to the charge carrier accumulation at the interface, the impedance gets a Warburg-like coefficient at high frequency and becomes purely capacitive at low frequency. On the other extreme, if the charge carriers are allowed to diffuse through the perovskite-metal electrode interface, then the contact becomes partially absorptive and there will be an additional semicircular arc at the lower frequency response of the impedance spectrum. In our present case, with the light exposure, the radius and height of the semicircular arc increase gradually becoming more resistive and less capacitive respectively implying a transition from photoconductive to photo resistive state.

The impedance spectroscopy was further utilized to investigate the frequency dependence of capacitance as shown in Figure 3c. We find that the high-frequency capacitive response does not change with light illumination with time. Previously, it is observed that high-frequency capacitance is usually dominated by geometrical and series resistance[58-59] and the present observation confirms that neither geometrical capacitance nor series resistance does play a major role here. On the other hand, the low-frequency capacitance is supposed to be increased rapidly under the photogeneration of carriers. However, in our case, the low-frequency response reduces upon 15 minutes of light exposure and regains back in dark within 180 minutes as also observed in the I-V study of these single crystals. Therefore, both the I-V characteristics and impedance measurement study confirm the existence of negative photo-conductivity (NPC) behavior in the as-grown $Cs_3Bi_2Cl_9$ crystal.

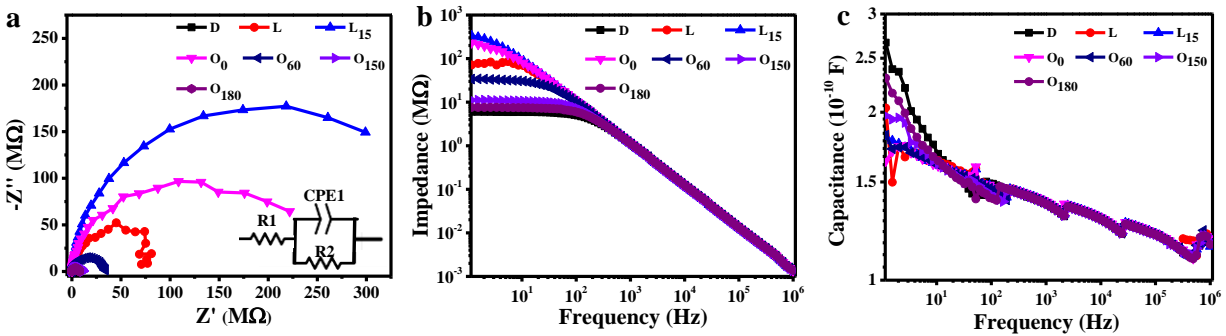

**Figure 3. Impedance spectroscopy characterization of $Cs_3Bi_2Cl_9$ single crystal.**
(a) The behavior of real and imaginary part of impedance (Nyquist plot) in the Dark, Light illumination, and again Light off condition. Inset: Equivalent circuit of this lead-free single crystal. In this circuit, resistance R1 is added in series and constant phase element CPE1 and resistor R2 are added in parallel. Here series resistance R1 is equivalent to contact resistance and R2 corresponds to bulk resistance of crystal. The constant phase element includes the bulk capacitance of the system.

(b) The behavior of the real part of impedance with Frequency in the Dark, Light illumination and again Light off condition.

(c) Capacitance vs frequency curve in the Dark, Light illumination and again Light off condition. In the low-frequency region, capacitance is decreasing in light illumination condition and again recovering in dark.

To understand the origin of the light-activated degradation and self-recovery behavior, photoluminescence spectra on single-crystal was performed in the dark and also after the exposure of 1-sun illumination for 15 minutes (Figure S3). However, no significant change in the PL position or intensity was observed which discards any light-induced structural or compositional changes. Furthermore, the role of ferroelectricity in the current degradation mechanism in the lead-free single crystal is eliminated in our study as observed timescales for current degradation and recovery are in minutes to hours. This timescale does not coordinate with the proposed domain migration time (milliseconds). Demonstration of a different underlying physical explanation of the photo-degradation/self-healing was also motivated by recently published reports which suggest the difference in the dielectric constant is associated with structural fluctuations, where photo-induced charge carriers change the polarizability.[60] The Raman measurement in dark and after 15 minutes of the light illumination (Figure S4) show no significant change, which eliminates the possibility of the local structural distortion and light-induced local distortions.[61] All these experimental results directly eliminate photo-induced structural changes, ferroelectricity or the local polarization effects as a plausible mechanism of the observed NPC phenomena.

To further explore the origin of observed NPC, we performed femtosecond transient reflectance (fs-TR) spectroscopy. The fs-TR measurements were carried out using ultrafast transient absorption (TA) spectrometer,[62] where the negative TR signal (*i.e.*, $\frac{R(t_i)-R(t_0)}{R(t_0)} = \frac{\Delta R}{R(t_0)} < 0$) and positive TR signal (*i.e.*, $\frac{R(t_i)-R(t_0)}{R(t_0)} = \frac{\Delta R}{R(t_0)} > 0$) are assigned to the photobleaching (PB) and photoinduced absorption (PIA), respectively. $R(t_0)$ and $R(t_i)$ are the reflectance signals without and with pumping at the delay time $t_i$. The FTRS analysis reveals the formation of the light-activated metastable trap states. Figure 4 shows the TR spectra of the $Cs_3Bi_2Cl_9$ SC after exciting at 350 nm (above band-edge). The negative reflectivity with maximum signal intensity at 383 nm refers to ground state bleach (GSB) observed due to the depletion of the charge carrier population. The GSB recovery kinetic (Figure 4b) depicts two-time constants, a fast 1.5 ps and a long 480 ps, which we attribute as trapping and charge carrier recombination time, respectively. However, at 415 nm excitation (just below the band-edge, Figure 4c), instead of GSB signal, only a broad PIA is noticed which can be referred to as trap state-assisted absorption. From TR spectra, it is clear that the intensity of trap states reflectivity close to bandgap is low whereas it gradually increases at low energy. The absence of a GSB signature on 415 nm excitation indicates no inherent mid-band state allows the two-photon absorption process.

To explore the nature of the trap state, we compare the TR kinetics at three different wavelengths (Figure 4d) that can be fit using a bi-exponential function with sub ps time scales. As can be seen from the kinetics, the excited state decay is the relatively slower rate in closed to bandgap (at 600 nm) than below bandgap (700 and 890 nm) states.

We further studied through exciting at 500 nm (below the band-edge) and the figures are summarized in Figure S6. Unlike 415 nm excitation, a relatively less broad and weak PIA is observed after 500 nm excitation. More importantly, the TR kinetics decay monitored in 735 nm to 800 nm in a similar way (Figure S6b). Furthermore, we compare the TR kinetics at 752 nm following 415 and 500 nm excitation (Figure S6c), which shows a significant difference in their decay rate. From the above discussion, it is obvious that the generation of trap states in $Cs_3Bi_2Cl_9$ SC is sensitive to light. The wavelength-dependent trap-assisted absorption can possible only when the trap states are generated through a light-induced process.

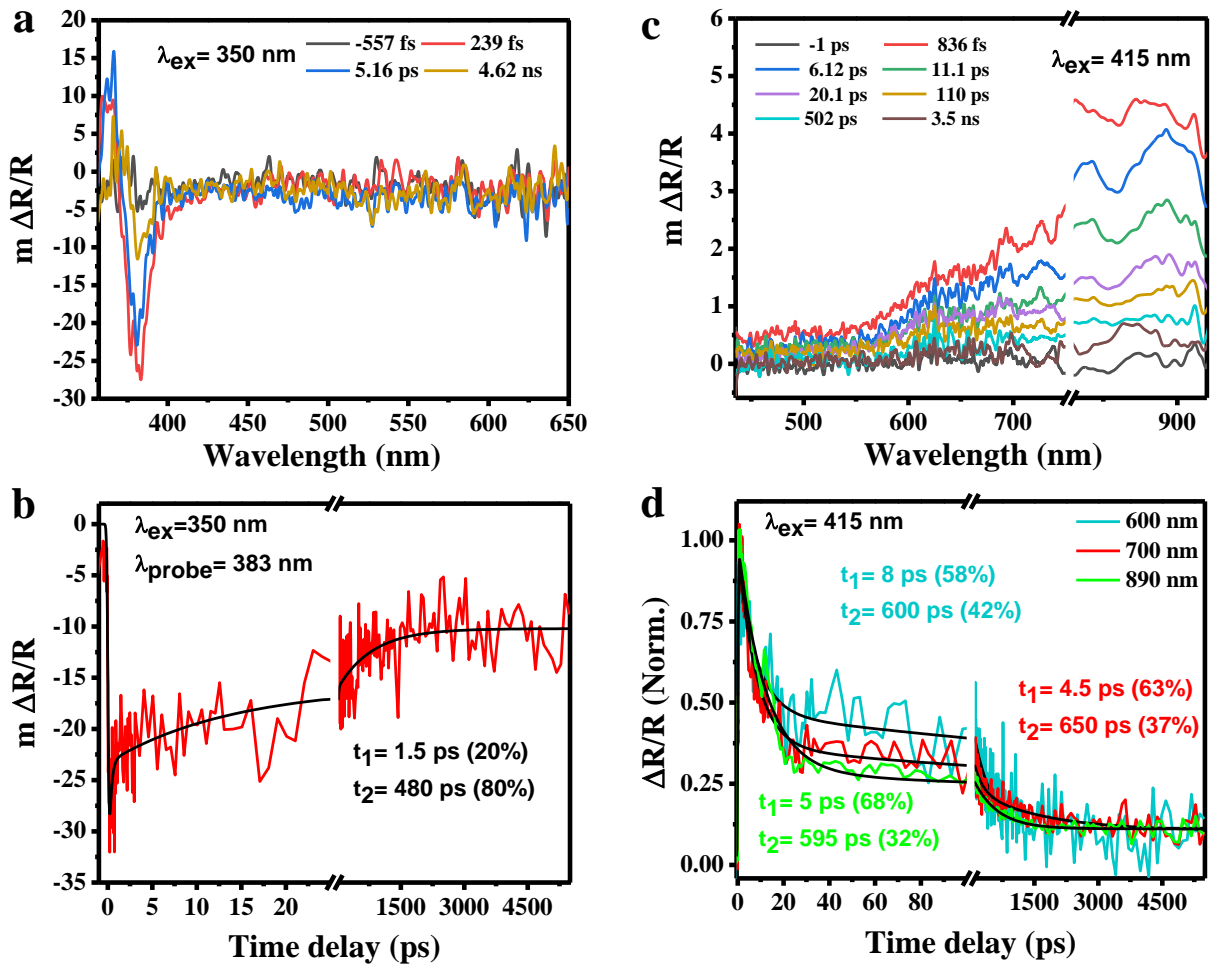

**Figure 4. Transient Reflection Spectrum**

(a,c) fs-TR spectra at a different time delay of $Cs_3Bi_2Cl_9$ SC in response to 350 and 415 nm excitation, respectively.

(b) GSB dynamic monitored at 383 nm after exciting at 350 nm.

(d) Normalized decay kinetics following 415 nm excitation. Solid black lines represent the exponential fitting of experimental data.

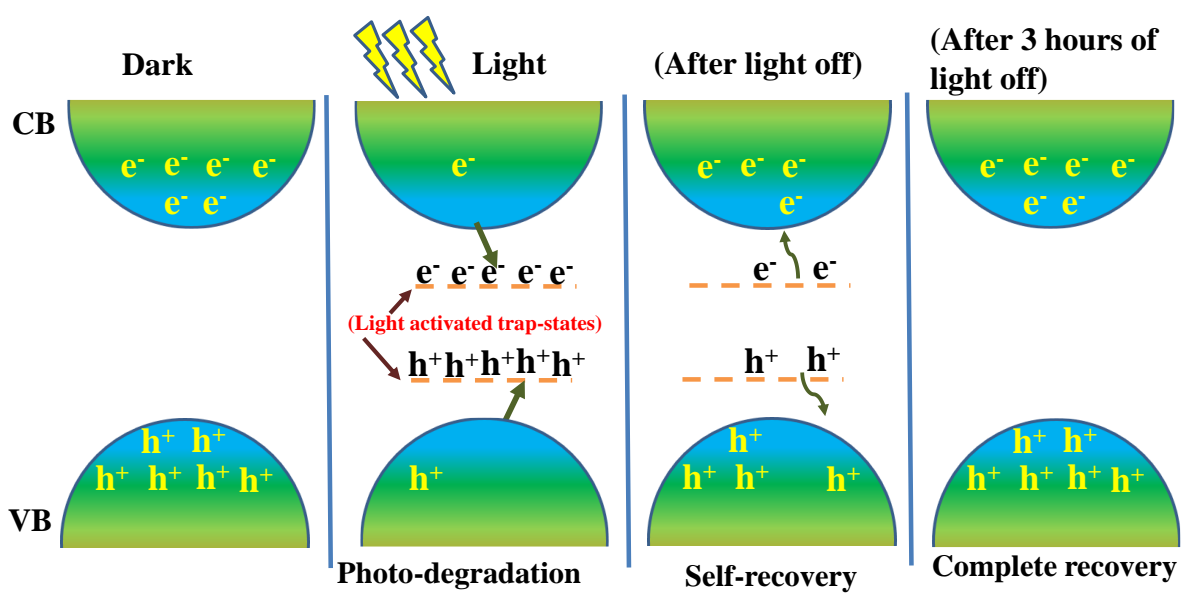

**Figure 5. Schematic Diagram of the Proposed Mechanism.**

Dark self-healing mediated negative photoconductivity mechanism in lead-free $Cs_3Bi_2Cl_9$ perovskite single crystal. Here diagram shows the evolution of the valence (VB) and conduction (CB) bands in three different situations. (i) During the dark condition (ii) during light-induced degradation and accumulation of charges and (iii) during recovery or self-healing in dark in dark. The dashed lines between bands represents the light-activated metastable trap states.

To demonstrate the observations in Figure 2-4, a mechanism is proposed based on the changes in the sub band-gap density of states over light-exposure duration in the $Cs_3Bi_2Cl_9$ perovskite single-crystal. As shown in Figure 5, broken red lines show an enhanced number of light-activated meta-stable states under constant light exposure. These meta-stable trap states accumulate over time evolution and form charged islands in the bulk of the single crystal, which, in turn, leads to the degradation of the observed dark current. On the other hand, when the device is kept in the dark, the majority of these light-activated trap states dissipate away, and self-healing occurs.

The observed negative photoconductivity phenomenon in $Cs_3Bi_2Cl_9$ single crystals can be used in the photodetection application with high responsivity and large detectivity. The responsivity and detectivity were calculated according to formula (SI) and values are estimated as 17 mA/W and $6.63 \times 10^{11}$ Jones at 5V bias voltage. We have observed enhancement in the responsivity and detectivity with the bias voltage as displayed in Figure S7a and S7b respectively. The ratio of

"current in dark" to "current after light incident 15 minutes" was estimated as ~7160 which is very higher in comparison to reported positive photodetectors. We observed that the responsivity and detectivity of the $Cs_3Bi_2Cl_9$ single crystals are comparable to other reported lead-halide perovskite single crystals based positive photodetectors.[29, 63-64]

**Conclusion**

In summary, the intrinsic photo-stability of lead-free perovskite $Cs_3Bi_2Cl_9$ single crystal is reported. From I-V and impedance measurements, this is concluded that these materials showed light-induced degradation and negative photoconductivity. From the transient reflectivity spectroscopy, it is assumed here light-illumination could induce meta-stable trap states generation which stated as responsible for this observed negative photoconductivity as well as the dark self-healing of photocurrent. This resistive switching behavior in perovskite single-crystal provides a deep understanding of the carrier trapping or de-trapping under the light on and off condition. These findings open up the need for further study about the interesting photophysical phenomena in these lead-free perovskite single crystals.

**Experimental Procedures-**

**Characterizations**

**Morphology:** Photographs of crystals were taken using a DSLR camera. A transmission microscope EVOS FLc was used for the surface image of crystals. The surface morphology images were taken using Field emission scanning electron microscope (FESEM) model FESEM QUANTA 200 FEG under electron excitation energy of 20 keV using ETD detector at 10000 magnification.

**Single crystal X-ray diffraction measurement:** Single-crystal X-ray diffraction data-sets of cesium bismuth chloride single crystal were collected using a Bruker Kappa APEX-II diffractometer equipped with a CCD detector using Mo-Kα radiation ($\lambda$ = 0.7107 Å). Data acquisition and processing were performed using the APEX2 software. The structure was analyzed and refined using the WinGX software package and Mercury software and Olex2 software.

**Optical measurements:** Ultraviolet-visible (UV-vis) absorption spectra of single-crystal were collected using a Microprocessor UV-Vis Double Beam spectrophotometer LI-2800. Photoluminescence measurements were performed with an RF-6000 Spectro-Fluorophotometer. For the Raman spectra measurement, the WITec alpha300 RA Raman instrument was used with the laser source 532 nm.

**TCSPC measurement:** For the lifetime measurement, a time-correlated single-photon counting (TCSPC) experiment was performed using FLOROCUBE set up and 334 nm wavelength NanoLED was used as an excitation source. Data were collected using the TBX-04D detector.

**Electrical measurement:** The current-voltage (I-V) characteristics have been measured using Integrated Photovoltaic Test System Assembly Instrument under standard AM1.5G conditions under dark and under illumination (100 mW/cm$^2$). IS measurements were recorded using Zahner Electrochemical Workstation in the dark and under illumination. Measurements were collected for a frequency range of 1 Hz and 1 MHz. The equivalent circuit was modeled using ZSimpWin 3.21 software.

**Femtosecond transient reflectance (fs-TR) spectroscopy:** fs-TR spectroscopy was performed at timescales from 0.1 ps to 5 ns in a transient absorption (TA) spectrometer at room temperature. In this study, an Ultrafast Helios fs-TA spectrometer equipped with UV-visible and near-infrared detectors was used. A few μJ/pulse energy of the amplified 800 nm laser beam (100-fs pulse width and a repetition rate of 1 kHz) was made to pass through a spectrally tunable (240-2600 nm) optical parametric amplifier (Light conversion) to generate the excitation pump pulses at 350, 415 and 500 nm. The probe pulses (UV visible and NIR wavelength continuum, white light) were created by passing another fraction of the 800 nm pulses through a 2-mm thick sapphire crystal. Before white light generation, the 800 nm pulses were passed through a motorized delay stage. Depending on the movement of the delay stage, the transient species were detected following excitation at different time scales. The white light was split into two beams (referred to as signal and reference) and focused on two fiber optics to improve the signal to noise ratio. The excitation pump pulses were spatially overlapped with the probe pulses on the samples after passing through a synchronized mechanical chopper (500 Hz), which blocked an alternative pump pulse. All spectra were averaged over a time of 2 s for each time delay. Absorption change (ΔA) was recorded by the Surface Xplorer software with respect to the delay time ($t_i$) and wavelength (λ, nm). Finally,

$\frac{\Delta R}{R(t_0)}$ was calculated using the equation of $\frac{\Delta R}{R(t_0)} = 10^{-\Delta A} - 1$. The $-\frac{\Delta R}{R(t_0)} < 0$ and $-\frac{\Delta R}{R(t_0)} > 0$ represents the photobleaching (PB) and photoinduced absorption (PIA), respectively, where R(t$_0$) and R(t$_i$) are the reflectance signals without and with pumping at the delay time t$_i$.

**Supplemental Information-**

The supplemental Information is attached separately. SEM image, Tauc plot, PL and Raman spectra before and after illumination, analysis of impedance data and fitting parameters, additional TA data, responsivity and detectivity graph.

**Data availability**

The data that support the plots within this paper and other findings of this study are available from the corresponding author upon reasonable request.

## ACKNOWLEDGEMENTS

NKT acknowledges the UGC Fellowship. SS acknowledges the Ministry of Electronics and Information Technology research grant DIC-1377-PHY. Authors also like to acknowledge Prof. Osman Bakr and Prof. Omar.Abdelsaboor from KAUST for helping in conducting transient reflection experiment and explain the data.

## AUTHOR CONTRIBUTIONS

Naveen Kumar Tailor has done the crystal synthesis, light induced photodegradation experiment and impedance spectroscopy experiment. Partha Maity has done and explained the transient reflection spectroscopy. Makhsud Saidaminov has helped in writing the manuscript and explaining the data. Narayan Pradhan has helped in writing the manuscript and explaining the data. Soumitra Satapathi has conceptualized the problem, explained the data and has helped in writing the manuscript.

## DECLARATION OF INTERESTS
Authors declare no competing financial interest.